\documentclass[%
 reprint,
superscriptaddress,
 amsmath,amssymb,
 aps,
 prx,
floatfix,
]{revtex4-2}

\usepackage{graphicx}
\usepackage{dcolumn}
\usepackage{bm}
\usepackage{siunitx}
\usepackage{xcolor}
\usepackage{csquotes}
\usepackage{nicefrac}
\usepackage{hyperref}
\hypersetup{
    colorlinks=true,
    linkcolor=blue,
    filecolor=magenta,      
    urlcolor=blue,
    citecolor=blue,
}
\usepackage{hyperref}

\begin{document}

\preprint{Fracional spin excitations in CaCuO$_2$}

\title{Fractional spin excitations in the infinite-layer cuprate CaCuO$_2$} 

\author{Leonardo Martinelli}
\email{leonardo.martinelli@polimi.it}
\affiliation{Dipartimento di Fisica, Politecnico di Milano, piazza Leonardo da Vinci 32, I-20133 Milano, Italy}
\author{Davide Betto}
\affiliation{ESRF, The European Synchrotron, 71 Avenue des Martyrs, CS 40220, F-38043 Grenoble, France}
\author{Kurt Kummer}
\affiliation{ESRF, The European Synchrotron, 71 Avenue des Martyrs, CS 40220, F-38043 Grenoble, France}
\author{Riccardo Arpaia}
\affiliation{Quantum Device Physics Laboratory, Department of Microtechnology and Nanoscience, Chalmers University of Technology, SE-41296 G\"oteborg, Sweden}
\author{Lucio Braicovich}
\affiliation{Dipartimento di Fisica, Politecnico di Milano, piazza Leonardo da Vinci 32, I-20133 Milano, Italy}
\affiliation{ESRF, The European Synchrotron, 71 Avenue des Martyrs, CS 40220, F-38043 Grenoble, France}
\author{Daniele Di Castro}
\affiliation{Dipartimento di Ingegneria Civile e Ingegneria Informatica, Università di Roma Tor Vergata, Via del Politecnico 1, I-00133 Roma, Italy}
\affiliation{CNR-SPIN, Università di Roma Tor Vergata, Via del Politecnico 1, I-00133 Roma, Italy}
\author{Nicholas B. Brookes}
\affiliation{ESRF, The European Synchrotron, 71 Avenue des Martyrs, CS 40220, F-38043 Grenoble, France}
\author{Marco Moretti Sala}
\affiliation{Dipartimento di Fisica, Politecnico di Milano, piazza Leonardo da Vinci 32, I-20133 Milano, Italy}
\author{Giacomo Ghiringhelli}
\email{giacomo.ghiringhelli@polimi.it}
\affiliation{Dipartimento di Fisica, Politecnico di Milano, piazza Leonardo da Vinci 32, I-20133 Milano, Italy}
\affiliation{CNR-SPIN, Dipartimento di Fisica, Politecnico di Milano, I-20133 Milano, Italy}

\date{\today}

\begin{abstract}
We use resonant inelastic x-ray scattering (RIXS) to investigate the magnetic dynamics of the infinite-layer cuprate CaCuO$_2$. We find that close to the $(\nicefrac{1}{2},0)$ point the single magnon decays into a broad continuum of excitations accounting for $\sim 80 \%$ of the total magnetic spectral weight. Polarization resolved RIXS spectra reveal the overwhelming dominance of spin-flip ($\Delta S = 1$) character of this continuum with respect to the $\Delta S = 0$ multimagnon contributions. Moreover, its incident-energy dependence is identical to that of the magnon, supporting a common physical origin. We propose that the continuum originates from the decay of the magnon into spinon pairs, and we relate it to the exceptionally high ring exchange $J_c \sim J_1$ of CaCuO$_2$. In the infinite layer cuprates long-range and multi-site hopping integrals are very important and amplify the 2D quantum magnetism effects in spite the 3D antiferromagnetic N\'eel order. 

\end{abstract}
%
%
\maketitle
%
\section{\label{sec:introduction} Introduction}
The spin $\nicefrac{1}{2}$ antiferromagnetic (AF) square lattice is one of the most studied systems in condensed matter theory, both for its exquisite quantum mechanical nature and for its implications in the longstanding problem of high $T_c$ superconductivity in cuprates \cite{keimer2015quantum}. Despite the apparent instability in 2D, the magnetic ground state of these systems has been experimentally determined to be of N\'eel–type mostly thanks to weak inter-layer exchange and Dzyaloshinskii-Moriya interactions \cite{christensen2007quantum, vaknin1987antiferromagnetism, endoh1988static}. Interestingly, the linear spin-wave theory (LSWT), which describes the magnetic excitations as spin-1 quasiparticles, provides an excellent description of the spin response in most of the 2D Brillouin zone \cite{christensen2007quantum, aeppli1989magnetic, coldea2001spin, peng2017influence}, and of other static thermodynamic properties \cite{chakravarty1989two, majumdar2012effects}. However, the region close to the magnetic zone boundary (related to short-wavelength physics) is known for deviating from this simple picture   \cite{christensen2007quantum,headings2010anomalous, dalla2015fractional, gretarsson2016persistent} with a sudden loss of magnon spectral weight and the build-up of a  continuum at higher energies in the vicinity of the $(\nicefrac{1}{2},0)$ point. These observations remain unexplained in LSWT, even when pushed to higher order of $\nicefrac{1}{S}$ expansion \cite{igarashi2012magnetic}. One possible interpretation is in terms of multimagnon processes, i.e., excitations involving a number of magnons greater than one  \cite{sandvik2001high, lorenzana1999does, powalski2018mutually}. And, going beyond LSWT, multimagnons have indeed been successfully used to partially reproduce  the experimental results, e.g. for
Sr$_2$CuO$_2$Cl$_2$ \cite{betto2021multiple}.
\begin{figure*}[t]
\includegraphics[width=0.8\textwidth]{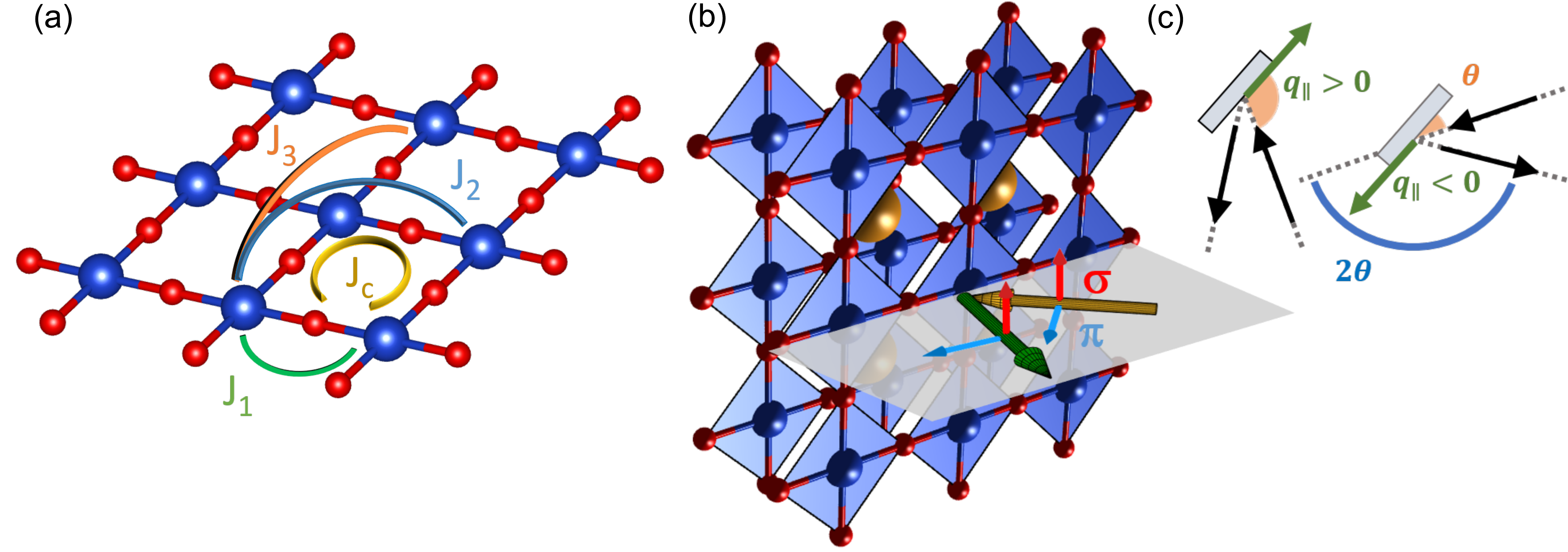}
\caption{\label{fig:geometry} (a) Scheme of CuO$_2$ planes, with Oxygen atoms depicted in red and Copper atoms in blue. Coloured lines highlight the magnetic coupling of the Hubbard-like Hamiltonian as explained in the text. (b) Sketch of CaCuO$_2$ crystal structure and experimental geometry used in the RIXS experiments, also showing the incident and outgoing polarization of light. Cu atoms are blue, oxygen atoms red and Ca cations are in gold. Polarization perpendicular to (lying in the) scattering plane is labelled $\sigma$ ($\pi$). (c) Scheme of grazing-in and grazing-out geometries. The shaded orange area represents the incident angle, the scattering angle $2\theta$ is depicted in blue, while the green arrows depict the component of the in-plane transferred momentum.}
\end{figure*}

However, several theoretical works have suggested that the peculiarities at the magnetic zone boundary arise from the proximity to some exotic non-magnetic phases, such as the resonating valence bond (RVB) \cite{dalla2015fractional, anderson1987resonating}, the AF* \cite{ho2001nature} or valence bond solid (VBS) \cite{sandvik2007evidence, tang2013confinement, shao2017nearly} states. The building block of these pure quantum mechanical ground states are pairs of spins arranged into local singlets \cite{anderson1987resonating}, and their most fascinating property is the fractionalization of magnons into unbound (or almost-unbound) magnetic excitations carrying spin \nicefrac{1}{2}, usually referred as spinons. This spinon band is gapped in the N\'eel ground state, with a dispersion minimum  at $(\nicefrac{1}{2},0)$ where the magnon energy is maximum \cite{shao2017nearly, dalla2015fractional, ho2001nature}. The mixing of the two bands at this wave vector might then transfer spectral weight from the magnon to the higher-energy spinon giving rise to the continuum measured around $(\nicefrac{1}{2},0)$ \cite{shao2017nearly}. Indeed, at short wavelengths, minimally deconfined spinon pairs become more similar to magnons and the two sorts of spin-1 excitations mix, similarly to an exciton-polariton scenario \cite{shao2017nearly}. Whereas in 1D the separation (deconfinement) of two spinons has been observed in the case of spin-$\nicefrac{1}{2}$ chains and ladders, in 2D deconfinment is predicted to be only partial \cite{shao2017nearly}, unless the system is a spin liquid \cite{tang2013confinement}. It has been proposed that such fractionalization might be present in the pure-Heisenberg AF compound copper deuteroformate tetradeurate, where the magnon anomaly has been clearly observed \cite{dalla2015fractional, shao2017nearly}. However, the experimental results were later accounted for also using LSWT with a strong magnon-magnon interaction \cite{powalski2018mutually}, casting doubts on the actual observation of spin fractionalization in that case. Therefore, to our knowledge, the existence of partially deconfined spinons in 2D square lattice has found no definitive experimental support so far.

Cuprates are obvious candidates for the observation of fractional magnetic excitations: although the RVB ground state originally proposed by Anderson \cite{anderson1987resonating} is challenged by the evidence of N\'eel AF order coming from neutron scattering and magnon dispersion, the $(\nicefrac{1}{2},0)$ anomaly is a common feature. The magnetic Hamiltonian of cuprates arises from a fundamental Hubbard-like physics \cite{zhang1988effective, ino1999fermi, damascelli2003angle, yoshida2006systematic} and entails couplings (schematized in Fig. \ref{fig:geometry}) among first ($J_1$), second ($J_2$) and third ($J_3$) nearest neighbours, plus another multi-spin term that couples four spins across a Cu$_4$O$_8$ plaquettes, usually called ring exchange \cite{peres2002spin, coldea2001spin, peng2017influence}. It has been put forward that the partial deconfinement of spinons occurs in the presence of  frustrating next-nearest neighbor  \cite{schulz1996magnetic,isaev2009hierarchical,capriotti2001resonating,ferrari2018spectral} or multi-spin couplings \cite{sandvik2007evidence,shao2017nearly,tang2013confinement,takahashi2020valence}. In particular, many studies have established that $J_c$ is by far the most important term \cite{lorenzana1999does, coldea2001spin, peng2017influence}. Among cuprates, the infinite-layer CaCuO$_2$ (CCO) is characterized by the strongest long range magnetic interactions, with a particularly large ring exchange $J_c$ \cite{braicovich2009dispersion, peng2017influence}, so that it is the ideal candidate for the quest of magnon fractionalization.

Although previous studies of CCO had already shown the presence of the anomaly and had brought an estimate of the $J$-values, highlighting the importance of the absence of apical oxygen in the structure \cite{peng2017influence}, the actual nature of the anomalous spin spectrum has till to be determined. Here we studied CaCuO$_2$ by resonant inelastic x-ray scattering (RIXS) at the Cu L$_3$ edge, exploiting several innovative techniques to unravel the properties of spin excitations at the $(\nicefrac{1}{2},0)$ point, and found evidence that this anomaly is more consistent with spinon pair continuum than with a multi-magnon picture.
Sec. \ref{sec:experimental} presents the technical experimental details of the measurements we have performed, including information about samples growth, while in Sec. \ref{sec:experimentalResults} we present the results of our measurements. In particular, Sec. \ref{sec:momentumDependence} reports the momentum dependence of the energy and intensity of the magnetic excitations, which highlight a clear anomaly close to $(1/2,0)$ point. In Sec. \ref{sec:polarimeter} we present the results of our polarimetric measurements, which favour an interpretation of the anomaly in terms of fractionalization over multiple magnons. Finally, in Sec. \ref{sec:detuning} we report the incident energy dependence of the magnetic spectrum which further corroborates our interpretation. Section \ref{sec:discussion} is devoted to the discussion and interpretation of the experimental results. 

\section{\label{sec:experimental} Experimental Details}
The crystal structure of CaCuO$_2$ displayed in Fig. \ref{fig:geometry}\textcolor{blue}{(a)}, is the archetype of infinite-layer cuprates, with a stack of CuO$_2$ planes separated by Ca$^{2+}$ cations.  Cu$^{2+}$ ions in the planes are in a 3$d^9$ configuration, with spin $\nicefrac{1}{2}$ and a single hole in the $3d$ shell. The absence of apical oxygen makes it almost impossible to dope this compound while preserving its special structure: doping by extra oxygen leads to a complex unit cell \cite{zhang1994identify}. Superconductivity has been obtained in heterostructures \cite{dicastro2015high} and superlattices \cite{Di_Castro_2014superconductivity, dicastro2012occurence} by
doping through the interface.

The CCO films were grown by pulsed laser deposition (KrF excimer laser, $\lambda = 248$ nm) at a temperature around 600 °C and an oxygen pressure of 0.1 mbar, on NdGaO$_3$ (NGO) (1 1 0) substrate. The substrate holder was at a distance of 2.5 cm from the CCO target, which was prepared by standard solid-state reaction \cite{Di_Castro_2014superconductivity, dicastro2015high}.
\begin{figure*}[t]
\includegraphics[width=\textwidth]{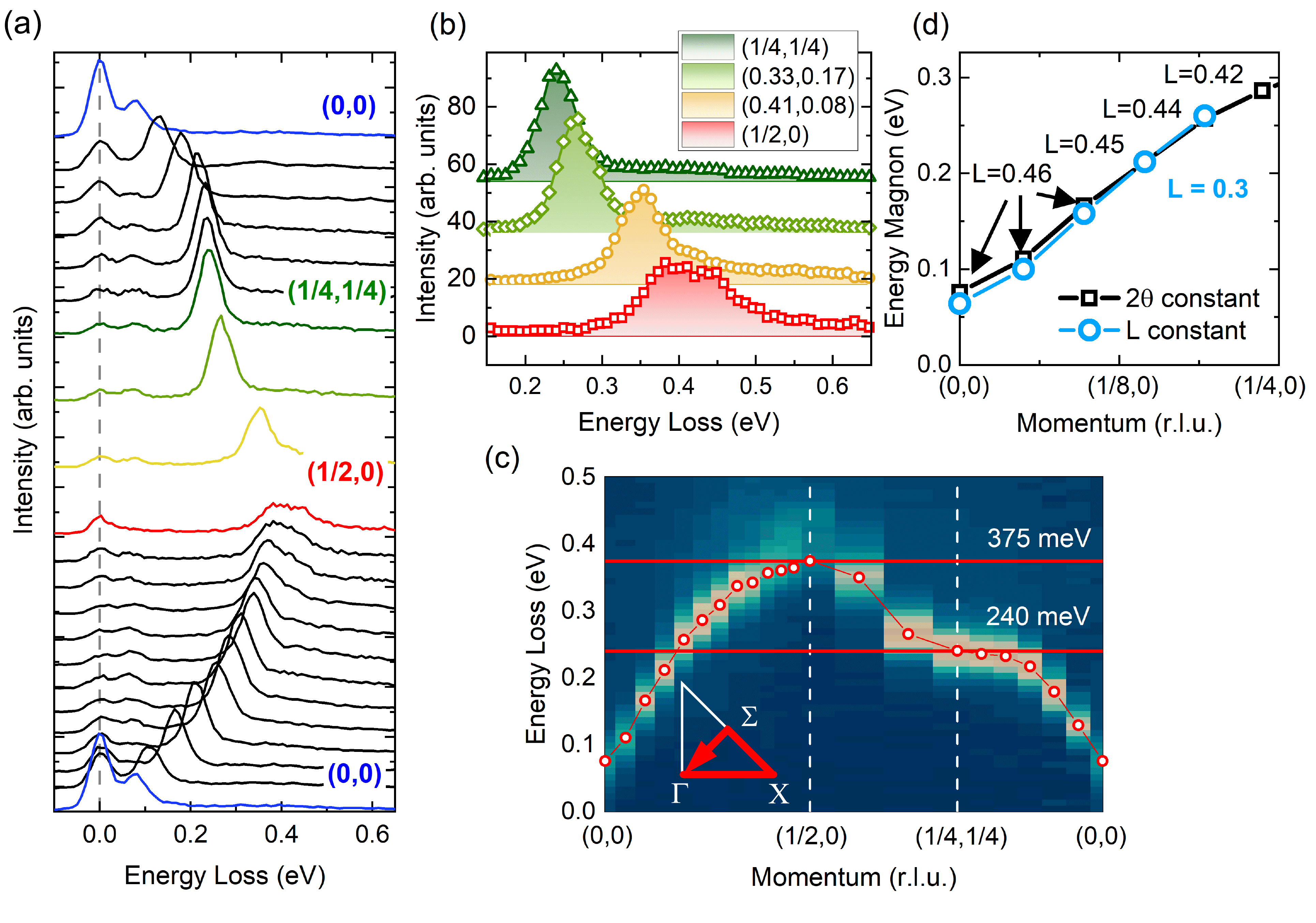}
\caption{\label{fig:overview} (a) RIXS scans at different momenta, acquired in grazing-out geometry and $\pi$ polarization. (b) Evolution of RIXS spectra along the $\Sigma \rightarrow X$ line. Scans have been normalized to spin-flip cross section.  (c) Measured dispersion of the magnon peak along the high-symmetry lines of the Brillouin zone illustrated in the inset. The red dots corresponds to the energy extracted by fitting the experimental data (error bars are smaller than the size of the dots). The two red horizontal lines highlight the magnon energy at ($\pi$,0) and at ($\pi$/2,$\pi$/2). (d) Energy of magnetic excitations along the $\Gamma \rightarrow X$ line measured with constant scattering angle (black squares) and constant $L$ (light blue circles). $L$ values at each point are reported. Incident polarization was $\pi$ and we employed grazing-out geometry.}
\end{figure*}

\par The RIXS measurements were performed at the beam line ID32 of the  European Synchrotron (European Synchrotron Radiation Facility, ESRF) in Grenoble (France), using the ERIXS spectrometer \cite{brookes2018beamline,braicovich2014simultaneous}. A schematic layout of the experimental geometry is depicted in Fig. \ref{fig:geometry} \textcolor{blue}{(b)}. The CuO$_2$ plaquettes were kept perpendicular to the scattering plane, and we denote $\sigma$ ($\pi$) the polarization perpendicular (lying into) the scattering plane, with the addition of a prime symbol ($\sigma', \pi'$) when referring to polarization of scattered photons.
To carefully investigate the magnetic spectrum we explored along all the degrees of freedom allowed by the RIXS technique: momentum transfer, incident-photon and scattered-photon linear polarization direction \cite{braicovich2014simultaneous}, and incident photon energy close to the absorption peak. The results are presented and discussed in sections \ref{sec:momentumDependence}, \ref{sec:polarimeter} and \ref{sec:detuning}, respectively. 

\par The momentum dependence was mainly performed by rotating the angle $\theta$ while keeping the scattering angle ($2\theta$) fixed at $\ang{149.5}$. We indicate the transferred momentum $\mathbf{q}_{\parallel}$ as $(H,K,L)$ in reciprocal lattice units (r.l.u.), i.e., fractions of the reciprocal lattice vectors $a'=2\pi/a$, $b'=2\pi/b$ and $c'=2\pi/c$. We conventionally assign positive (negative) values of $H$ and $K$  to grazing-out (grazing-in) geometries as depicted in Fig. \ref{fig:geometry}\textcolor{blue}{(c)}. Moreover we indicate the high-symmetry in-plane points $(0,0)$, $(\nicefrac{1}{2},0)$ and $(\nicefrac{1}{4},\nicefrac{1}{4})$ as $\Gamma$, X and $\Sigma$, respectively.
We also acquired some scans at fixed $L$ value, exploiting the continuous rotation of the ERIXS spectrometer to change the angle $2\theta$ \cite{brookes2018beamline}. The resolution was $\sim 48$\,meV unless differently indicated. In order to determine precisely the shape of the magnetic excitations, we also collected one high-resolution ($\sim 26$ meV) spectrum at $(-0.45,0)$, in a grazing-in configuration and with $\sigma$ incident polarization. The incident energy was fixed at the resonance peak of the Cu L$_3$ absorption edge ($\sim 931$\,eV), and the counting time was 30\,min for the medium resolution measurements and 5\,h for the high-resolution one.

Polarimetric spectra of section \ref{sec:polarimeter} were acquired at $(+\nicefrac{1}{2},0)$ using both $\sigma$ and $\pi$ incident polarization with a resolution of $45$\,meV, better than in previously published measurements \cite{fumagalli2019polarization,fumagalli2020mobile,hepting2018three}. Both scans have been acquired for 6\,h to compensate for the $\sim 10\%$ efficiency of the polarization analyzer. The error-bars have been estimated using the formulas given in Ref.~\onlinecite{fumagalli2019polarization}. In this geometry we obtained the parallel- ($\pi \pi', \sigma \sigma'$) and the crossed- ($\pi \sigma', \sigma \pi'$) polarization spectra corresponding to spin-conserving ($\Delta S =0$) and spin-flip ($\Delta S = 1$) final states respectively. The former can be assigned to two-magnon excitations, the latter to the magnon and related fractional continuum, and to odd number of multi-magnons. This type of information is thus crucial to determine the character of the continuum at the X point. 

Detuned RIXS spectra of section \ref{sec:detuning} were acquired at three incident photon energies below the XAS resonant peak ($\Delta E_{\mathrm{in}}=-0.150, -0.450, -0.675$\,eV), and performed at the $(-0.43, 0)$ point, i.e., in a grazing-in configuration, with $\sigma$ incident polarization. In this configuration we could maximize the count rate and have comparable cross sections for the  $\Delta S = 0$ and $\Delta S=1$ transition channels, and thus distinguish the nature of the various spectral features as being akin to the magnon or the two-magnon. In fact the intensity scales with the absorption for the former while it falls more rapidly for the latter, as discussed in references \onlinecite{rossi2019experimental} and \onlinecite{braicovich2020determining}.  Spectra at increasing distance from the resonance peak have been acquired for longer times to compensate for the signal decrease, from 30\,min to around 2\,h.  All the measurements were performed at 20\,K to minimize radiation damage.

\section{\label{sec:experimentalResults} Experimental Results}

\subsection{\label{sec:momentumDependence} Momentum dependence}
The scans at different $\mathbf{q}_\parallel$ are shown in Fig. \ref{fig:overview}\textcolor{blue}{(a)}.
The $\pi$ polarization in grazing-out configuration  enhances the magnetic excitations irrespective of the outgoing polarization \cite{sala2011energy,ament2011resonant}. Below we will often refer to the RIXS intensity in this geometry as purely magnetic, i.e., we will assume it to be proportional to the dynamical spin susceptibility, because the charge response is zero in this energy range in the undoped material, and we will neglect the small $\pi \pi'$ spectral contribution. Far from the $(\nicefrac{1}{2},0)$ point, a very sharp magnon peak dominates the spectra, on top of a very weak continuum. The energy width of the magnon, correlated with its lifetime, is very small in most of the Brillouin zone and even resolution-limited close to the $(\nicefrac{1}{4},\nicefrac{1}{4})$ point. This is due to the absence of doping and to the very high quality of the sample (long range AF order, little damping by magnetic disorder). A clear reduction in the magnon weight is seen when moving towards the X point.  At the same time, another structure extending from the magnon energy up to $\sim 800$\,meV is evidently increasing in intensity. While this behavior had already been observed in cuprates before \cite{coldea2001spin,igarashi2012magnetic,peng2017influence}, we underline that in the other cuprates the single magnon is anyway the dominant excitation with respect to the continuum  (see e.g. \cite{betto2021multiple}). On the contrary, in CaCuO$_2$, the single magnon is submerged by a broad and asymmetric continuum and becomes almost undetectable. Panels \ref{fig:overview}\textcolor{blue}{(b,c)} clearly show that this anomalous behaviour is restricted to a region of radius $\sim 0.15$ r.l.u. around the X point.
%
%
%
\begin{figure}[ht]
\includegraphics[width=0.48\textwidth]{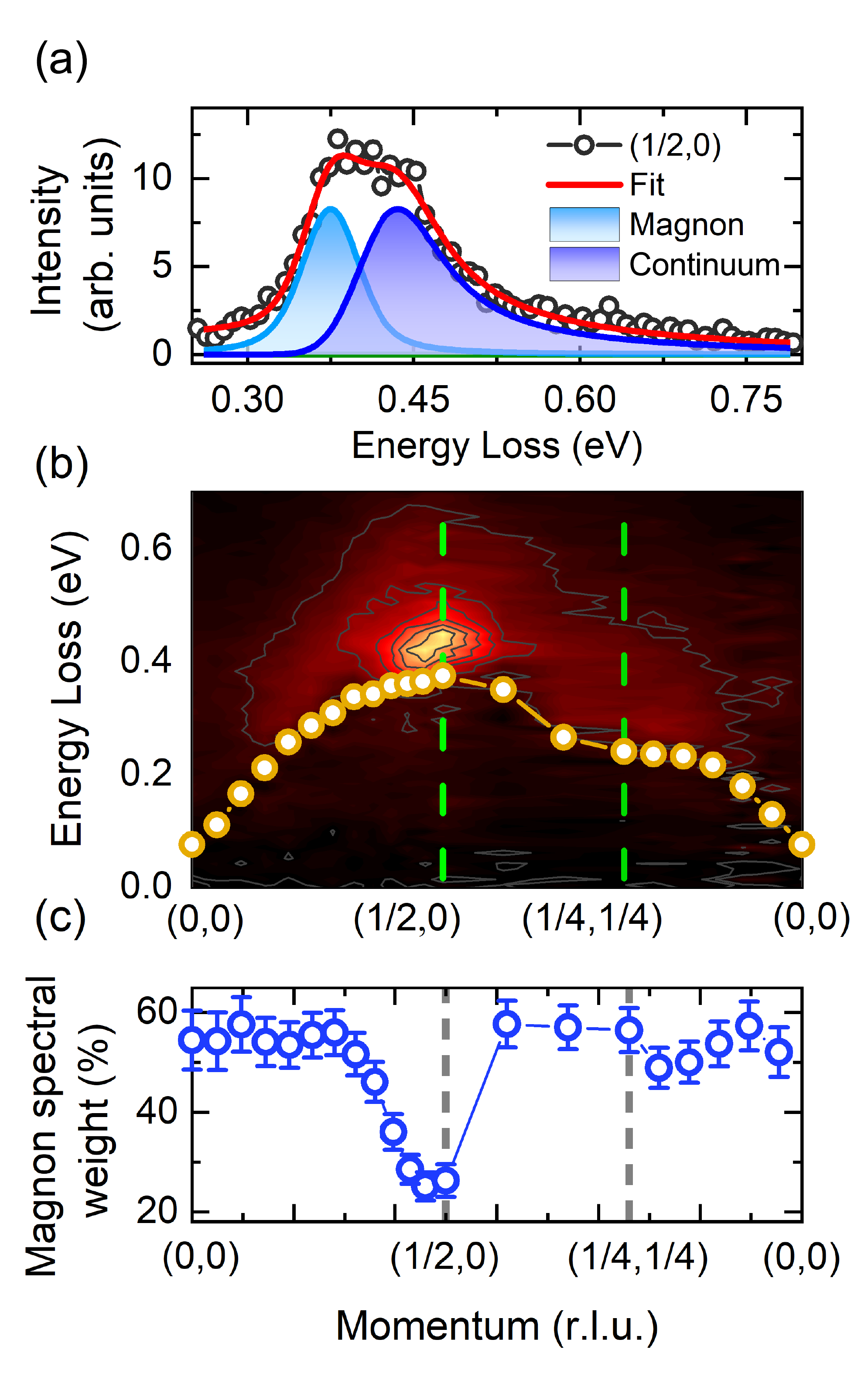}
\caption{\label{fig:anomaly} (a) Example of fitting procedure of the spectrum at $(0.5,0)$ collected with $\pi$ polarization. In the legend, we indicate for clarity only the magnetic features (magnon peak in light blue and continuum in deep blue) and the resulting fitted curve (in red). (b) Momentum dependence of high-energy continuum. Elastic, phonon and magnon peaks have been removed for clarity. The energy of the magnon is reported in yellow dots. For clarity, the scans have been normalized to the atomic $\pi\sigma'$ RIXS cross-section, which slightly modulates the intensity but only depends on experimental geometry \cite{sala2011energy}. (c) Ratio between the weight of magnon peak, as extracted from the fitting, and the total weight of the magnetic excitations.}
\end{figure}
\begin{figure*}[t]
\includegraphics[width=0.9\textwidth]{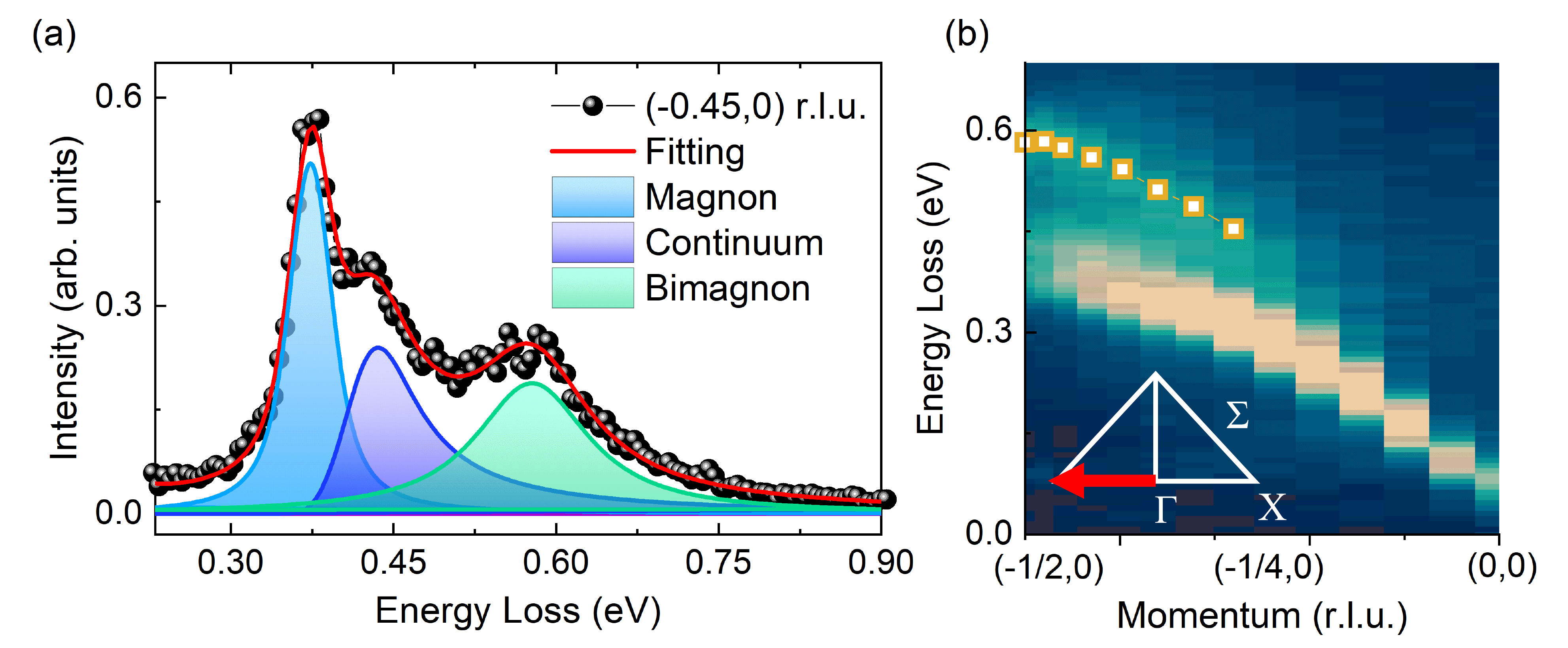}
\caption{\label{fig:highRes} (a) High resolution ($\approx 26.3 $ meV) spectra acquired at (-0.45,0) using $\sigma$ polarization. The use of high resolution clearly allows to disentangle the single magnon peak and the bimagnon peak from the high-energy tail. (b) Map of RIXS spectra along the $\Gamma \rightarrow$ (-X) line, acquired with $\sigma$ incident polarization and grazing-in geometry. Momentum values are reported with positive sign for clarity. The energy of bimagnon excitations extracted with a fitting is highlighted in yellow squares.} 
\end{figure*}

As mentioned above, scanning $\mathbf{q}_\parallel$ at constant $2\theta$ implies that the out-of-plane component of the transferred momentum changes too. In particular, $L$ ranges from $\sim 0.46$ to $\sim 0.22$ when moving from $(0,0)$ to $(0.5,0)$. Since CCO is known to possess a more 3-dimensional magnetic structure as compared to the other cuprates, implying a non-zero energy of the magnon at $(0,0,L)$ with non-integer value of $L$  \cite{peng2017influence, andersen1996out}, we also acquired some RIXS spectra along a limited range of the same in-plane path but at constant $L=0.3$. While the  magnon energy is different at the $\Gamma$ point, the difference decreases at larger in plane momenta and vanishes for $H>\nicefrac{1}{8}$, as shown in Fig.~\ref{fig:overview}\textcolor{blue}{(d)}. We thus assume that the anomaly at the X point is negligibly influenced by the value of $L$.

\par The false-color map of Fig.~\ref{fig:overview}\textcolor{blue}{(d)} shows the RIXS  intensity along the in-plane momentum path depicted in the inset, with the energy of the single magnon marked by the red dots. The large energy $E_\text{X}\approx 375$\,meV at $(\nicefrac{1}{2},0)$ and the strong dispersion along the magnetic zone boundary ($\Delta E_{\text{MZB}} = E_X - E_\Sigma \approx135$\,meV) are signatures of the fact that $J_1$ and $J_c$ are both very large, probably the largest among all layered cuprates \cite{peng2017influence,plumb2014high}.

%
%

As the magnon dispersion was previously analyzed in detail \cite{peng2017influence}, we concentrate here on the anomaly at the X point, with the aim of determining if the single magnon dominating everywhere else is replaced here by multiples or by fractions of magnons. To extract the exact momentum-dependent intensity of this continuum, we have fitted the spectra in the $[0,0.8]$\,eV energy range as depicted in Fig.~\ref{fig:anomaly}\textcolor{blue}{(a)}: a Gaussian shape for the the elastic/quasi-elastic and phonon peaks (assumed to be resolution limited), a Voigt profile for the single magnon (whose intrinsic line width might not be negligible), and an asymmetric Lorentzian profile for the high-energy continuum above it; spectral weights have been taken as the area below the curves. In Fig.~\ref{fig:anomaly}\textcolor{blue}{(b)} we plot the intensity map along the $\Gamma - $X$ - \Sigma - \Gamma$ line after removal of the elastic, phonon and single magnon contributions; the energy of the magnon is indicated by the yellow dots. 


The results of the fitting of our RIXS spectra allow us to quantify the single magnon spectral weight decrease at the X point by normalizing it to the total RIXS intensity in the fitted energy range, which is predominantly coming from the $\Delta S = 1$ magnetic excitations. The result is shown in Fig. \ref{fig:anomaly}\textcolor{blue}{(c)}. Away from the $(\nicefrac{1}{2},0)$ point, the single magnon contribution is quite constant and makes up for roughly 60\% of the total RIXS intensity, while this value drops by a factor of 3 to $\sim 20 \%$ at the X point. This deviation seems to be confined within $0.15$ r.l.u. around X.

\begin{figure}
\includegraphics[width=0.48\textwidth]{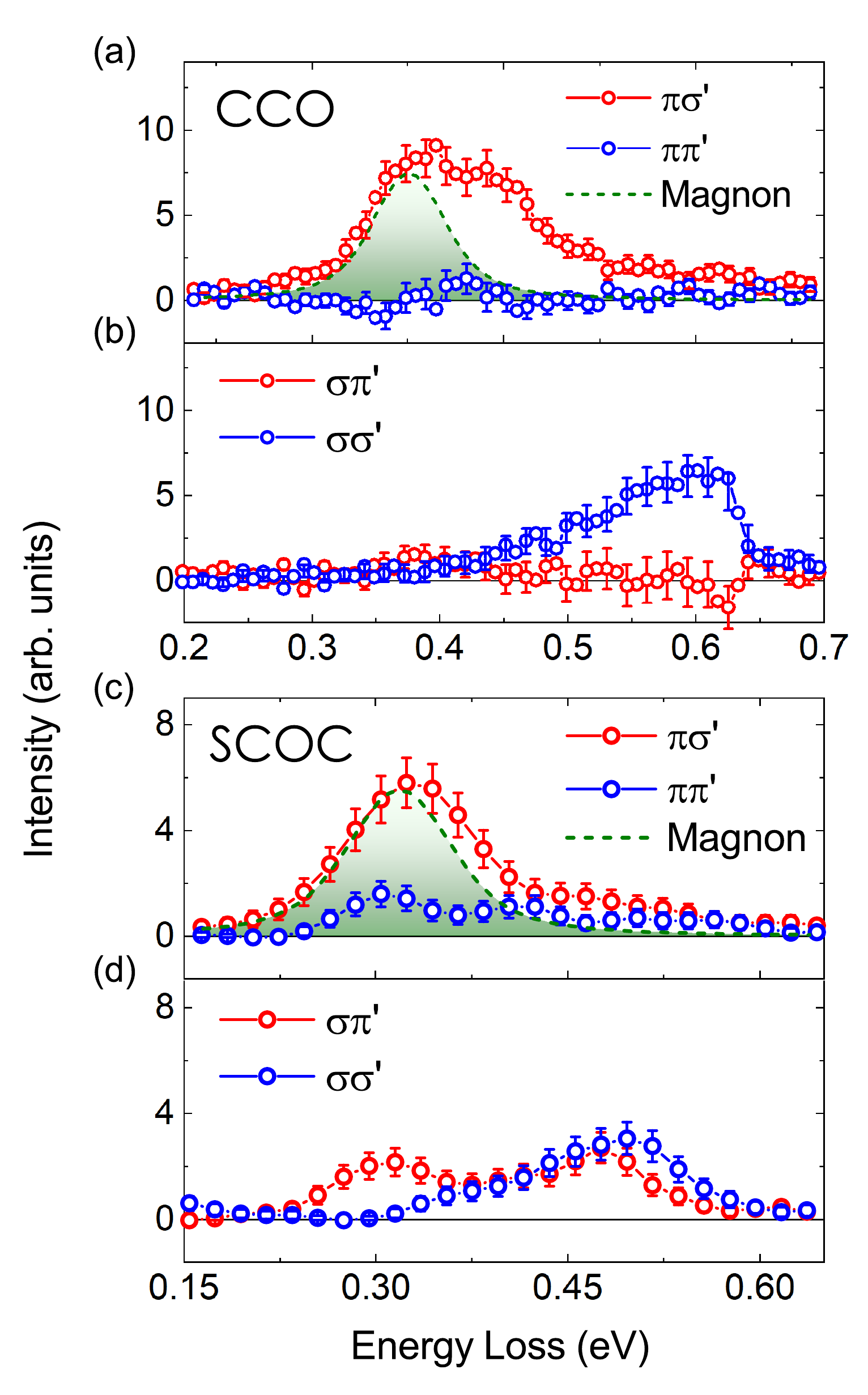}
\caption{\label{fig:polarimeter} Polarimetric spectra collected at $(\nicefrac{1}{2},0)$ for CCO (left panels) and SCOC (right panels). Panels (a) and (c) show the spectra acquired at ($+\nicefrac{1}{2}$,0) with incident $\pi$ polarization, while panels (b) and (d) show the spectra measured with incident $\sigma$ polarization to maximize the signal coming from non-spin flip excitations. Blue curves always denote the parallel channel, while red ones the crossed channel. Note that the resolution of the SCOC data is worse ($\approx 65$ meV) \cite{betto2021multiple}, so that the lineshape appears to be broader.}
\end{figure}
The anomalous shape of the magnetic spectrum close to the X point might be seen as a strongly damped (i.e., intrinsically broad) magnon \cite{le2011intense, minola2015collective, peng2018dispersion}, possibly combined with a \enquote{local} bi-magnon excitation of comparable intensity \cite{ament2011resonant,braicovich2009dispersion}. To check for the amount of bi-magnon contribution to the continuum, we have performed measurements with $\sigma$-polarized incident photons and a grazing-in geometry (negative $H$), for which the single-ion cross sections predict appreciable intensity for both  $\sigma\sigma'$ and $\sigma\pi'$ channels \cite{sala2011energy}, i.e., the $\Delta S = 0$ (even order multiple magnon) and $\Delta S = 1$ (single magnon) contributions, respectively. We exploited the best available energy resolution at the beam line ($26$\,meV) to define as precisely as possible the intrinsic shape of the magnetic spectrum at the $(-0.45,0)$ point, where the magnon peak is still discernible while a sizable continuum is already visible. The spectrum of  Fig.~\ref{fig:highRes}\textcolor{blue}{(a)} shows that the high energy continuum is not a tail of a broad magnon but rather an independent well defined spectral feature. Fitting the spectrum we obtain an almost resolution-limited magnon at 340\,meV and a satellite peaked at 440\,meV with a width of $33\pm 5$\,meV. The use of $\sigma$ polarization also allows us to detect a well-defined feature around 540\,meV, whose incident-energy and polarization dependence (see below) identify it as a local bi-magnon
\cite{sala2011energy, ament2011resonant}. The momentum-dependence along the $\Gamma \rightarrow $ X line shown in Fig.~\ref{fig:highRes}\textcolor{blue}{(b)} confirms that this feature is more intense closer to X and disperses upward as predicted by calculations for bi-magnons \cite{ament2011resonant,igarashi2012magnetic}. We can therefore exclude that the continuum closer to the magnon peak is a local bi-magnon. This implies that also in the spectra measured with $\pi$ polarization at positive $H$, a configuration that anyway minimizes the even multiple-magnon contribution, the continuum is not due to local bi-magnons. 

\subsection{\label{sec:polarimeter} Polarimetric measurements}
We have shown that the continuum is not due to the relatively sharp bi-magnon observed for example in S$_2$CuO$_2$Cl$_2$ (SCOC) \cite{betto2021multiple}, but this does not rule out that it might be due to other types of spin-conserving excitations, i.e., an incoherent continuum of two magnons.
To truly determine its nature, we performed RIXS measurements with polarization analysis on the scattered beam \cite{braicovich2014simultaneous,fumagalli2019polarization,brookes2018beamline}.
%
%
By disentangling the vertical and horizontal polarization components of the scattered X-rays we were able to separate the crossed and parallel channels, corresponding to excitations with $\Delta S = 1$ and  $\Delta S = 0$, respectively. Indeed, although significant, the distinction of the two channels based on the incident photon polarization and on the RIXS cross sections in the single-ion model is not always conclusive. In fact, it was demonstrated that for SCOC non-local scattering operators have to be introduced to fully account for the RIXS spectral shape close to the X point \cite{betto2021multiple}.

\par For convenience we present in Fig.~\ref{fig:polarimeter} the complete set of polarization-resolved spectra of CCO and SCOC measured at the X point. In both samples the $\pi$ spectra are dominated by the crossed polarization components (red symbols) but, whereas the parallel (blue) contribution is non-negligible in SCOC, it is practically zero everywhere in CCO. Therefore also the high-energy satellite of the magnon of CCO clearly belongs to the $\Delta S=1$ channel. On the contrary, when exciting with $\sigma$ polarization the spectra are predominantly of parallel character (blue) in CCO and of mixed character in SCOC. Hence, the relatively sharp peak around $540$ meV $\approx 3J$, clearly resolved in the high-resolution spectra of CCO, can be regarded as a local bi-magnon excitation. Interestingly, in the large-$J_c$ CCO we observe no signature of the complex multi-magnon dynamics emerging in the small-$J_c$ SCOC \cite{betto2021multiple}.
\subsection{\label{sec:detuning} Energy dependence of magnon and continuum}
\begin{figure}
\includegraphics[width=0.48\textwidth]{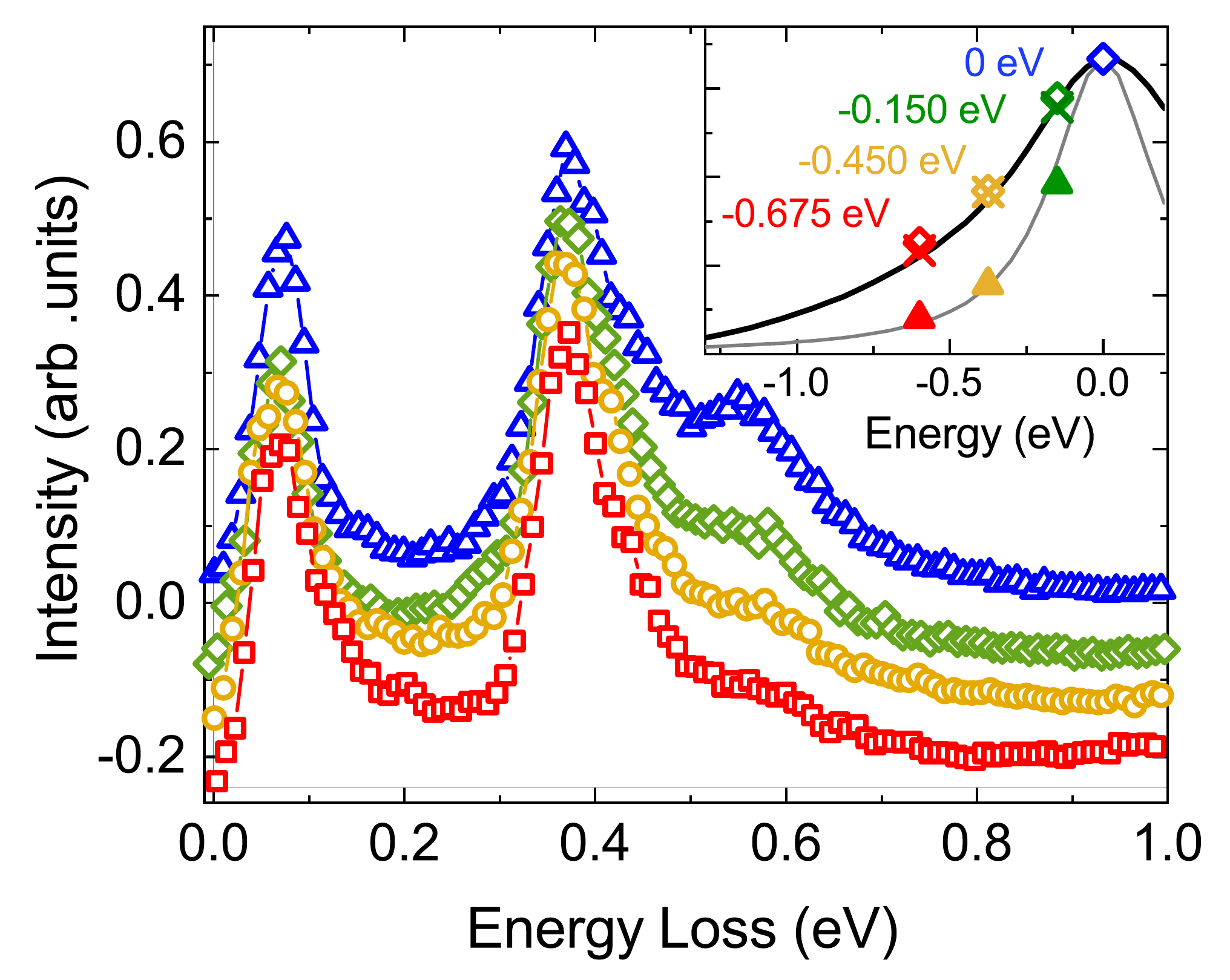}
\caption{\label{fig:detuning} Inelastic RIXS spectra at different incident energies. The elastic line has been removed for clarity. All the spectra were acquired at (-0.43,0) using $\sigma$ polarization, in order to have appreciable intensity both for the magnon and bi-magnon signal. The scans have also been normalized to the value of the XAS.
The inset shows the XAS signal (solid black line), the intensity of the magnon peak (empty diamonds), of the continuum (crosses) and of the bimagnon (solid triangles). The gray solid line is a guide to the eye. All intensities have been normalized to the resonance value, and the energies are referred to the maximum of the XAS (931.6 eV).}
\end{figure}

To further investigate the nature of the magnetic excitations we acquired RIXS measurements at different incident photon energies (so-called \enquote{detuning} analysis).
Fig. \ref{fig:detuning} depicts some RIXS scans at selected incident energies. By fitting the spectra we estimate the relative weights of the magnon, the continuum and the bi-magnon excitations. By comparing their incident energy dependence to the XAS profile (see inset) we observe that the $\Delta S = 1$ structure above 400 meV share the same energy dependence of the single magnon and of the XAS, as opposed to the bi-magnon intensity that decreases faster upon detuning, as previously observed \cite{bisogni2014femtosecond, braicovich2020determining,rossi2019experimental}. This is a further evidence that the continuum gets excited via the same process as the magnon and its intensity is not related to the lifetime of the RIXS intermediate state, which is not the case for the bi-magnon.
%

\section{\label{sec:discussion} Discussion}

We have found a strong experimental evidence that the continuum appearing in the RIXS spectra when approaching the X point has $\Delta S = 1$ character, thus compatible with being the decay of magnons into spinon pairs \cite{sandvik2007evidence,dalla2015fractional,ferrari2018spectral,shao2017nearly}. 
As fractional excitations in spin $\nicefrac{1}{2}$ square lattices have been predicted in several Heisenberg-like Hamiltonians \cite{schulz1996magnetic, isaev2009hierarchical, ho2001nature, capriotti2001resonating, sandvik2007evidence, dalla2015fractional, shao2017nearly,takahashi2020valence}
spinon pairs have been often invoked to explain the anomalous high-energy dynamics of cuprates \cite{headings2010anomalous} and isostructural iridates \cite{gretarsson2016persistent}. However, an undisputed experimental evidence of their existence in 2D is still missing. It is important to highlight that spinon pairs are predicted also in the presence of N\'eel antiferromagnetic order in the square lattice. Indeed, calculations predict that the proximity of the N\'eel phase to some exotic magnetic phase gives rise to a broad, gapped spinon band above the magnon energy, dispersing as  \cite{ho2001nature, suwa2016level, ferrari2018spectral}
\begin{equation}
  E(H,K) \approx \sqrt{E_0^2-(E_1 \cos(2\pi H) \cos(2\pi K))^2}.
  \label{spinpair_disp}
\end{equation}
The spinon pair dispersion reaches a maximum at $(\nicefrac{1}{4},\nicefrac{1}{4})$ and a minimum at $(\nicefrac{1}{2},0)$. Here, where the magnons reach their maximum energy, the mixing of spinon pairs with magnons becomes possible. The transfer of spectral weight is made possible by the quasi degeneracy in energy and by the common $\Delta S = 1$ nature of magnons and spinon pairs \cite{haldane1991fractional}). It is important to note that two-spinon excitation dispersion in 1D spin systems has been observed by RIXS \cite{schlappa2012spin,fumagalli2020mobile}. 


To understand why the X-point anomaly in the magnetic spectrum is larger in CCO than in other cuprates, we look at the magnetic exchange constants. We consider the usual spin Hamiltonian, which can also be obtained as fourth order expansion of the one-band Hubbard model  \cite{coldea2001spin, peres2002spin, delannoy2009low}, based on the definitions of Fig. \ref{fig:geometry}\textcolor{blue}{(a)}
\begin{equation}
\begin{split}
    H = &\ J_1 \sum_{ii'} \mathbf{S}_i \cdot \mathbf{S}_{i'} + J_2 \sum_{ii''} \mathbf{S}_i \cdot \mathbf{S}_{i''} + J_3 \sum_{ii'''} \mathbf{S}_i \cdot \mathbf{S}_{i'''} \\   
        &+ J_c \sum_{\langle ijkl \rangle} (\mathbf{S}_i \cdot \mathbf{S}_j)(\mathbf{S}_k \cdot  \mathbf{S}_l)+(\mathbf{S}_i \cdot \mathbf{S}_l)(\mathbf{S}_k \cdot \mathbf{S}_j) \\
        &\qquad \qquad - (\mathbf{S}_i \cdot \mathbf{S}_k)(\mathbf{S}_j \cdot \mathbf{S}_l)
    \end{split}
    \label{eqn:ehm}
\end{equation}
%
A simple analytical calculation relates the magnon energy at $(\nicefrac{1}{2},0 )$ and $(\nicefrac{1}{4},\nicefrac{1}{4})$ and the values of $J_1$ and $J_c$ \cite{coldea2001spin,delannoy2009low}:
\begin{equation}
\label{eqn:jc}
   \frac{J_1}{J_\text{c}} = \frac{3}{10} \frac{1+\Delta E_\text{MZB}/E_\text{X}}{\Delta E_\text{MZB}/E_\text{X}} 
\end{equation}
%
In the case of CCO, with the values extracted from Figure \ref{fig:overview}\textcolor{blue}{(c)}, equation \textcolor{black}{(\ref{eqn:jc})} translates in $J_c \approx J_1$. This value of $J_\text{c}/J_1$ is the largest among cuprates: for example, in the RBa$_2$Cu$_3$O$_6$ family this ratio is $0.8$, in La$_2$CuO$_4$ $\sim 0.29$, in single-layer Bi$_2$Sr$_2$CuO$_6$ $\sim 0.62$ \cite{peng2017influence} and in Sr$_2$CuO$_2$Cl$_2$ $\sim 0.42$ \cite{plumb2014high,betto2021multiple}. We note, incidentally, that the value $J_c/J_1$ determined here is slightly different from the ones reported in references \onlinecite{peng2017influence,braicovich2009dispersion}: the increased resolution allowed us to determine the shape of the spectrum more precisely, and to correctly identify the position of the single magnon even close to $(\nicefrac{1}{2},0)$. 

It is therefore natural to correlate the high-energy continuum, which we interpret as made of fractionalized magnetic excitations, to the strong $J_\text{c}$ which in this material cannot be regarded as a small perturbation. The presence of spinons in the spectrum of the Heisenberg AF with strong multi-spin couplings (of which the ring exchange is an example) has been predicted some years ago and is now well established \cite{sandvik2007evidence, tang2013confinement, shao2017nearly, takahashi2020valence}. Interestingly, a recent exact diagonalization study \cite{larsen2019exact} performed on the EHM Hamiltonian has shown that, at $J_c/J \approx 1$ the excitation spectrum at $(\nicefrac{1}{2},0)$ breaks into a continuum of states very close in energy, similarly to what has been calculated for Neél-RVB and Neél-VBS transitions \cite{sandvik2007evidence, schulz1996magnetic, ferrari2018spectral,isaev2009hierarchical,capriotti2001resonating}.  


\section{\label{sec:conclusion} Conclusions}
We have performed a deep investigation of the magnetic excitations in the infinite-layer cuprate CaCuO$_2$ using RIXS, analysing their dependence on momentum, incident and outgoing polarization, and incident energy. The momentum dependence reveals a clear anomaly, close to the $X$ point in reciprocal space, which is clearly not explainable in terms of a simple magnon with increased width. The incident polarization dependence rules out that this anomaly can be due to local bimagnon excitations. Conversely, its interpretation in terms of a continuum of spinon pairs is strongly supported both by the polarimetric analysis and by the incident energy dependence, which clearly shows that this anomalous scattering channel originates from a direct RIXS process.
A deviation from the LSWT scenario is a general property of all square-lattice AF materials, even of systems where the nearest-neighbour coupling is the only dominant interaction \cite{christensen2007quantum}. However, the measured continuum is greatly enhanced in this material even with respect to other cuprates, and at the same time the multi-magnon features seem to be suppressed. An estimate of the next nearest neighbor magnetic  interactions suggests that the cause of these peculiarities lies in the exceptionally large value of ring exchange coupling $J_c$, and in general in large values of the superexchange integrals, having its root in the absence of the hole-localizing potential provided by the apical oxygens \cite{peng2017influence}. It would therefore be of interest to perform the same experimental investigation on   Nd$_2$CuO$_4$, in which the absence of ``direct'' apical oxygens \cite{skanthakumar1989magnetic} should enhance the ring exchange \cite{peng2017influence}, similarly to what happens in CCO. 

In conclusion, we provide here the experimental evidence of fractional spin excitations in the magnetic spectrum of a layered cuprate. Spinon pairs are a well established feature of 1D magnetic systems and were predicted by different theoretical approaches to be present also in 2D AF lattices \cite{sandvik2007evidence, shao2017nearly, tang2013confinement, ferrari2018spectral, ho2001nature}. However, their observation in layered cuprates had remained somehow not conclusive so far \cite{headings2010anomalous} and our results can support the assignment of the magnon anomaly close to the X point to fractional spin excitations also in other cuprates. More in general, we demonstrate that multi-spin couplings play a non-marginal role in determining the short-wavelength physics of the magnetic excitations, as indeed proposed by several theories \cite{sandvik2001high, tang2013confinement, larsen2019exact}.

\section*{Acknowledgments}
We acknowledge José  Lorenzana for enlightening discussions.
The RIXS experimental data were collected at beamline ID32 of the European Synchrotron (ESRF) in Grenoble, France, using the ERIXS spectrometer designed jointly by the ESRF and Politecnico di Milano. 
M. M. S. and G. G. acknowledge support by the project PRIN2017 “Quantum-2D” ID 2017Z8TS5B of the Ministry for University and Research (MIUR) of Italy. 
R.A. acknowledges support  by the Swedish Research Council (VR) under the Project 2020-04945.


\bibliography{spinons}

\end{document}